
\input phyzzx
\def\mgut{M_{\rm GUT}}
\def\mpl{M_{\rm Planck}}
\def\ggut{g_{\rm GUT}}
\def\gstr{g_{\rm string}}
\def\msb{{\overline{\rm DR}}}
\def\Str{\mathop{\rm Str}\nolimits}
\def\str{\mathop{\rm str}\nolimits}
\def\wf{{\cal W}^{\rm field}}
\def\ws{{\cal W}^{\rm string}}
\def\b{{\bf B}}
\def\B{{\cal B}}
\def\eqdef{\buildrel\rm def\over{\Relbar\joinrel\Relbar}}
\interdisplaylinepenalty=10000
\advance\normaldisplayskip by 0pt plus 2pt minus 2pt
\Twelvepoint
 \Pubnum={Stanford--ITP--838}
 \date={December 1987}
\Pubtype={Revised May 1992}
\titlepage
\title{One-Loop Threshold Effects \penalty-1000
    in String Unification\foot{%
         Research supported in part by the NSF under grant
         \#~PHY--86--12280.}}
\author{Vadim S.\ Kaplunovsky\foot{%
    Current (1992) address: Dept.\ of Physics, University of Texas,
    Austin, TX~78712.
    Current research is supported in part by the NSF under grant
    \#PHY--90--09850 and by the Robert~A.\ Welch Foundation.}}
\address{Department of Physics\break Stanford University\break
    Stanford, CA 94305}

\abstract
Like grand unification of old, string unification predicts simple
tree-level relations between the
couplings of all unbroken gauge groups
such as $SU(3)_C$ or $SU(2)_W\)$.
I show here how to compute one-loop corrections to these
relations for any four-dimensional model based on a
classical vacuum of the heterotic string.
The result can be used to calculate both $\sin^2\theta_W$ and
$\Lambda_{\rm QCD}$ in terms of $\alpha_{\rm QED}$ and $\mpl\)$.

\par\nobreak\vfil\nobreak\medskip
   \centerline{Printed (with minor errors) in
       \sl Nuclear Physics \bf B307 \rm (1988)}
\endpage
 \sequentialequations
\chapter{Introduction}
The term {\it grand unification\/} usually refers to unification
of low-energy gauge interactions into a single gauge theory such as
$SU(5)$, $SO(10)$ or $E_6$ that is spontaneously broken at some
high energy scale $\mgut\sim10^{14}$--$10^{19}$~GeV.
Not surprisingly, the main subjects of grand unified theories (GUTs)
are the low-energy gauge couplings and the Georgi-Quinn-Weinberg
(GQW) equations
\Ref\GQWref{H.~Georgi, H.~R.~Quinn and S.~Weinberg
 \journal Phys. Rev. Lett. &33 (74) 451.}
that relate these couplings to $\mgut$ and to each other.
In the first approximation GQW equations can be written as
$$
  \forall a:\quad {16\pi^2\over g_a^2(\mu)}\
  =\ k_a\cdot{16\pi^2\over \ggut^2}\
  +\ b_a\cdot\log{\mgut^2\over\mu^2}\ +\ O(1)\,,
 \eqn\GQWnaive
$$
where index $a$ runs over low-energy gauge couplings and
$\mu$ is some phenomenological scale such as $M_W\)$.
Here the coefficients $k_a$ are the tree-level relations between
the couplings (\eg\ $k_1=5/3$, $k_2=k_3=1$ in $SU(5)$
\Ref\GG{H.~Georgi and S.~L.~Glashow \journal Phys. Rev. Lett.
 &32 (74) 438}%
) while $b_a$ are the effects of couplings' renormalization below
$\mgut$: they are related to the one-loop $\beta$-functions
via $\beta_a=b_a\cdot g_a^3/16\pi^2$.
For a given grand unified model both $b_a$ and $k_a$ are fixed
numbers
and there are only two free parameters:
$\ggut$ and $\mgut\)$.
Hence, GQW equations  yield one parameterless relation between
the three phenomenologically known couplings $g_1\)$, $g_2$ and $g_3$
--- or between related quantities $\alpha_{\rm QED}\)$,
$\sin^2\theta_W$ and $\log\Lambda_{\rm QCD}$;
this relation is often called a prediction for $\sin^2\theta_W\)$.
In addition we can also compute $\mgut$ which is important for
predicting the rate of proton decay and other processes that involve
the unified theory.

The accuracy of equations \GQWnaive\ allows one to compute
$\sin^2\theta_W$ up to terms of order $O(\alpha)$ and $\log\mgut$
up to terms of order $O(1)$, which means that only the order of
magnitude of $\mgut$ can be reliably predicted.
In order to increase the accuracy of these predictions one has to
solve
the two-loop renormalization group equations for the couplings that
are relevant below the GUT scale,
and the boundary conditions for those equations should include
one-loop corrections to the tree-level relations between the
coupling.
At the one-loop level of accuracy
gauge couplings of the effective subthreshold-energy theory
\Ref\SW{S.~Weinberg \journal Phys. Lett. &91B (80) 51.}
generally look like
$$
  \forall a:\quad {16\pi^2\over g_a^2(\mu)}\
  =\ k_a\cdot{16\pi^2\over\ggut^2}\
  +\ b_a\cdot\log{\mgut^2\over\mu^2}\ +\ \Delta_a \,.
 \eqn\deltaeq
$$
The new elements here are the explicit
$O(1)$ {\it threshold corrections} $\Delta_a$ whose values depend
on masses of heavy particles
associated with the unification threshold.

{\it String unification\/} is more ambitious than ordinary grand
unification: superstring theories
\Ref\stringrev{For review see {\it ``Superstring Theory''\/}
 by M.~B.~Green, J.~H.~Schwarz and E.~Witten,
 Cambridge University Press (1987).}
are unified theories of {\it all\/} particle interactions.
All currently known semi-realistic string models are based on the
heterotic string theory,
\Ref\GHMR{D.~J.~Gross,
 J.~A.~Harvey, E.~Martinec and R.~Rohm \journal Phys.\ Rev.\ Lett.
 & 54 (85) 502 \journal Nucl.\ Phys. & B256 (85) 253
 \journal Nucl.\ Phys. & B267 (86) 75.}
so this article deals exclusively with four-dimensional models of
this kind.
As far as low-energy gauge interactions are concerned,
tree-level predictions of the string unification
are similar to the GUT predictions:
$(g_a^{\rm tree})^{-2}=k_a/\gstr^2\)$,
\Ref\PG{P.~Ginsparg \journal Phys. Lett. & 197B (87) 139.}
so the GQW equations \GQWnaive\ apply equally well for both kinds
of unification.
Other tree-level couplings are also proportional to powers of
$\gstr$;
in particular, the Newton's constant is given by
$G_N=\alpha'\gstr^2/8\pi$.
Moreover, $\mgut^2=O(1/\alpha')$,
\Ref\VSK{V.~Kaplunovsky\journal Phys. Rev. Lett. &55 (85) 1036.}
so $\mgut/\gstr=O(\mpl)$, which eliminates one free parameter from
the GQW equations~\GQWnaive\ and make
both $\sin^2\theta_W$ and $\Lambda_{\rm QCD}$ predictable
in terms of $\alpha_{\rm QED}$ and $\mpl\)$.
Again, in order to make those predictions accurate
one has to solve the two-loop renormalization group equations
and to compute the threshold corrections~$\Delta_a$
for the string threshold;
this time, one also has to know the relation between $\mgut$
(as used in eqs.~\deltaeq) and $\alpha'$ or $\mpl\)$.
\foot{Actually we only need to know the $O(1)$ term in
 $\log\left({\mgut\over\mpl}\right)$.
 Hence, we need not worry about string loops correcting the
tree-level
 relation between $\alpha'$, $\gstr$ and $\mpl\)$.
 For the same reason, an $O(1)$ term correction to $1/\gstr^2$
 would also be unimportant.
}

The goal of this article is to derive a general formula for threshold
effects in superstring theory.
Specifically, I shall compute $\Delta_a$ for any
four-dimensional model that is
based on a classical vacuum of the heterotic string
and has no tachyons;
the model need not be space-time supersymmetric.
This article is organized as follows:
In the next section I derive
a formula for threshold corrections in ordinary GUTs;
the result is not new, but the procedure provides a basis for the
subsequent string calculation.
The string calculation is presented in section~3;
a general formula relating threshold effects in a string model to
the model's massive spectrum concludes that section.
In the final section~4 I show how to apply the general formula to
orbifold models and carry out the calculations for the $Z_3$ model.

\chapter{Thresholds in Ordinary Gauge Theories}

I would like to begin with an outline of a general procedure for
calculating threshold effects.
When we want to compute \eg\ $\Delta_{\rm QCD}\)$ in some GUT,
we first select a process that is dominated by QCD
at energies below the unification scale
--- for example, gluon scattering.
Second, we compute the one-loop amplitudes for this process
as given by the two theories:
\par\begingroup\parskip=0pt
\item{A)} Complete unified theory that accounts for all particles,
light and heavy.
\item{B)} Effective low-energy theory, namely QCD with quarks
 and other light particles, but without any particles
 of mass $O(\mgut)$ or heavier.
\par\noindent\endgroup
At this point we do not use renormalization group techniques, but
instead provide both theories with explicit ultraviolet regulators.
The specifics of regularization schemes are not important as long as
the ultraviolet regulators do not affect the infrared behavior of the
respective theories; notice that this
requirement excludes the dimensional regularization, but
allows such schemes as Pauli-Villars, higher derivatives, proper-time
cutoff, lattice, \etc{}

If the theory $B$ is the effective low-energy theory describing
the same world as the more fundamental theory $A$, then the two
amplitudes should agree with each other at low energies $E$, \ie
$$
  {\cal A}_A(g_A,\Lambda_A,E,\ldots)\ =\
  {\cal A}_B(g_B,\Lambda_B,E,\ldots)\ +\ O\bigl(E^2/\mgut^2\bigr)
 \eqn\amplitude
$$
(I assume both cutoffs $\Lambda_A$ and $\Lambda_B$ are chosen to be
high enough to safely ignore any term proportional to a
negative power of $\Lambda_{A,B}$).
For given cutoffs $\Lambda_{A,B}\)$, equation \amplitude\ yields a
relation between the two bare couplings $g_A$ and $g_B\)$.
Having derived this relation, we can now use the
renormalization group equations to replace the bare coupling $g_B$
with
the running QCD coupling $g_{\rm QCD}(\mu)$;
if the amplitude ${\cal A}_B$ was computed correctly, the cutoff
$\Lambda_B$ should disappear from the resulting formula.
Similarly, $g_A$ and $\Lambda_B$ can be replaced with the running
unified coupling $\ggut(\mu')$.
It remains to substitute $\mu=\mu'=\mgut\)$, and we have a threshold
relation between $\ggut$ and $g_{\rm QCD}$;
other low-energy gauge couplings can be related to $\ggut$ in the
same
way.
Note that the definition of a running coupling $g(\mu)$ depends on a
choice of a {\sl renormalization scheme}.
Hence the values of the one-loop threshold corrections $\Delta_a$
depend on a choice of renormalization schemes for both $g_a$ and
$\ggut$;
for the gauge couplings, I shall use the $\msb$ scheme
(modified minimal subtraction for dimensional reduction)
throughout this article.

The above procedure is general enough to be valid in any unified
theory, be it an ordinary GUT or a superstring-based model.
However, when it comes to choosing a process, the state-of-the-art
string theory imposes a limitation: only
on-shell amplitudes can be computed at present.
Fortunately, an on-shell amplitude does not have to be a
scattering amplitude:
Lagrangian density of a non-trivial background
that obeys equations of motions is also on-shell.
\foot{This is indeed fortunate since the loop corrections to on-shell
scattering amplitudes have bad infrared problems in any interacting
theory with massless particles.
In string theory those problems become much worse
because we can no longer separate one-particle irreducible
Feynman diagrams from corrections to external legs.}
The simplest background for probing the gauge coupling $g_a$ is given
by a uniform (\ie, covariantly constant) gauge field $F^a_{\mu\nu}$;
the effective Lagrangian density of this background
has been used before for comparing unlike ultraviolet
regulators for QCD such as lattice versus Pauli-Villars,
\Ref\HH{A.~Hasenfratz and P.~Hasenfratz
 \journal Phys. Lett. & 93B (80) 165.}
and I simply generalize this approach to the case when
a unified theory acts as a QCD cutoff.

For any background that solves the classical equations of motion,
quantum field theory gives the following expression for the one-loop
contribution to its Lagrangian density:
$$
  {\cal L}_1\ =\ \half\,\Str\left(\,\log L\,\right)\ \equiv
  \int_0^\infty\!{dt\over2t}\;
  C_\Lambda(t)\cdot\Str\bigl(e^{-tL}\bigr) ,
 \eqn\propertimeq
$$
where the super-trace is taken over the Hilbert space of
all quantum fields (including ghosts).
Operator $L$ is the first-quantized Lagrangian:
$L=M^2-D^\lambda D_\lambda$ for scalar and ghost fields,
$L=M^2-D^\lambda D_\lambda -\half F_{\mu\nu}\gamma_\mu\gamma_\nu$
for spinors ($L$ is the second order Lagrangian)
and $(L)_{\mu\nu}=M^2-D^\lambda D_\lambda -2F_{\mu\nu}$
for vectors in the background Feynman gauge
($D_\mu$ are covariant derivatives with respect to the
background gauge potentials $A_\mu$).
The integration variable  $t$ is the proper time
and $C_\Lambda(t)$ is an ultraviolet regulator.

The bare coupling $1/g_a^2$ is the coefficient of the
$\coeff14 {F^a_{\mu\nu}}^2$ in the bare Lagrangian,
so what we need is the coefficient ${\cal W}_a$ of a similar
term in ${\cal L}_1\)$.
Let me therefore expand the exponential in \propertimeq\ to the
second
order in $A_\mu$ and $F_{\mu\nu}\)$,
substitute $F_{\mu\nu}(x)=\rm const$, $A_\mu(x)={-1\over2}
F_{\mu\nu}x^\nu$ and do the $x$ integral (or the equivalent
momentum integral) that is implicit in the super-trace.
The result is
$$
  \wf_a \ =\ {1\over 16\pi^2} \int_0^\infty\! {dt\over t}\;
  C_\Lambda(t)\cdot\left[ \b_a(t)\,\eqdef\,2\str \left(
  Q_a^2\,(\coeff{1}{12}-\chi^2)\,e^{-tM^2}\right)\right] ,
 \eqn\fieldren
$$
where the super-trace is now taken over the spectrum of
physical particles of the theory,
operator $Q_a$ is a generator of the gauge-group~$a$,
$\chi$ is the helicity operator and $M$ is the mass operator.
Notice that the integral in \fieldren\ diverges in the infrared
$t\to\infty$ limit;
the divergence is logarithmic, with a coefficient related to
the $\beta$-function of the low-energy theory:
$$
  \b_a(t)\,\becomes{t\to\infty}\,
  -\coeff{11}{3}\tr_V(Q_a^2) \,+\, \coeff23\tr_F(Q_a^2)
  \,+\, \coeff13\tr_S(Q_a^2)\ \equiv\ b_a
 \eqn\irfield
$$
where the traces are taken over the massless charged particles
{\sl and count each CPT-conjugate particle-antiparticle pair only
once}.

Formula \fieldren\ is applicable to any kind of a gauge theory.
The effective low-energy theory has the same massless spectrum
as the complete unified theory  but no massive particles.
Hence, while for the complete theory $\b_a(t)$ is $t$-dependent,
for the subthreshold theory, $\b^{\rm eff}_a(t)\equiv b_a\)$.
For both theories $(g_a^{\rm bare})^{-2}+{\cal W}_a$ should have
the same value, so
$$
 \eqalign{
  {16\pi^2\over g_a^2}\,-\,{16\pi^2 k_a\over\ggut^2}\
  &= \int_0^\infty\!{dt\over t}\,\left( C_{\rm GUT}(t)\cdot\b_a(t)
                            - C_{\rm eff}(t)\cdot b_a\right)\cr
  &= 2\hbox{\fourteenrm str}_{M\neq0}
      \left( Q_a^2 (\coeff1{12} -\chi^2 ) \,
            \log{\Lambda_{\rm PV}^2\over M^2}\right) ,\cr}
 \eqn\fielddiff
$$
where I have computed the integral for the case when the same
Pauli-Villars regulator $C_{\rm PV}(t)=1-e^{-t\Lambda^2}$
is used for both theories.
It remains to translate the bare couplings on the left hand side of
\fielddiff\ into the running $\msb$ couplings evaluated at
$\mu=\mgut$;
the result is the following expression for
the {\it threshold corrections in grand unified models\/}:
$$
  \Delta_a\ =\ 2\hbox{\fourteenrm str}_{M=O(\mgut)}
  \left( Q_a^2 (\coeff{1}{12}-\chi^2 ) \log{\mgut^2\over M^2}\right)
{}.
 \eqn\ordinary
$$

I would like to conclude this section with a remark that
formula~\ordinary\ applies not just to the GUT threshold, but to
any threshold that separates two ordinary gauge theories.
This includes any intermediate-energy thresholds a string-based model
might possess:
once we have an effective four-dimensional
field theory at some sub-Planck energy scale,
its further evolution towards lower energies
can be handled with ordinary field-theoretical techniques.
It is the string threshold itself that requires a different formula;
I shall derive it in the next section.

\chapter{One-Loop Effects at the String Threshold}

Having established a general procedure for computing one-loop
threshold effects,
let me now apply it to the string threshold.
My starting point is the string analogue of formula~\propertimeq:
$$
  {\cal L}^{\rm string}_1\ = \int_\Gamma {d^2\tau \over \tau_2}\cdot
  \left[\,Z(\tau,\bar\tau)\,\eqdef\,
  \Tr\left(q^H\,\bar q^{\bar H}\right)\,\right] ,
 \eqn\vacener
$$
where $Z(\tau,\bar\tau)$ is the partition function of the string,
$H\eqdef L_0-{c\over24}$ and $\bar H\eqdef \bar L_0-{\bar c\over24}$
are the left-moving and the right-moving world-sheet Hamiltonians,
$g\eqdef e^{2\pi i\tau}$, and $\Gamma\eqdef
\{\tau=\tau_1+i\tau_2\,:\, \tau_2>0,|\tau_1|<\half,|\tau|>1\}$
is the modular domain of the one-loop world sheet.
Formula \vacener\ was derived by Polchinski
\Ref\JP{J.~Polchinski\journal Comm. Math. Phys. &104 (86) 37.}
for the closed bosonic string in 26 flat space-time dimensions,
or in any other background that obeys the classical equations of
motion;
such backgrounds correspond to
conformally invariant field theories on the world sheet.
For the heterotic string one should replace a single
partition function $Z$ with a sum of separate partition functions
for the four spin structures:
$$
  Z(\tau,\bar\tau)\,\mapsto\,\half\sum_{s_1,s_2}
  \left[\, Z_{s_1s_2}(\tau,\bar\tau)\ \eqdef\ \Tr_{s_1}
  \left( (-)^{s_2\cdot F}\, q^H\,\bar q^{\bar H}\right) \right] ,
 \eqn\heterz
$$
where traces are now taken over distinct Ramond ($s_1=1$) and
Neveu-Schwarz ($s_1=0$) sectors of the Hilbert space.

Let us chose a classical vacuum of the heterotic string
that gives rise to a four-dimensional model with some massless
particles such as gauge bosons.
When some space-time fields $\phi$ acquire vacuum expectation values,
this amounts to a perturbative change in the string's background.
As long as a perturbed background remains a solution to classical
equations of motion, formula~\vacener\ will remain applicable.
The partition function~$Z_\phi$ of the new background is
perturbatively related to the original background via
$$
  Z_\phi\ =\ Z_{\phi=0} \cdot \left( 1\,+
  \int\!\!d^2\zeta\vev{V_\phi(\zeta,\bar\zeta)}\,+
  \,\half\!\int\!\!\!\!\int\!\!d^2\zeta_1\,d^2\zeta_2
  \vev{V_\phi(\zeta_1,\bar\zeta_1)\, V_\phi(\zeta_2,\bar\zeta_2)}\,
  +\,\cdots\,\right) ,\!
 \eqn\perturbation
$$
where $\zeta$'s are coordinates on the world torus,
$V_\phi(\zeta,\bar\zeta)$ are string vertices corresponding
to the background fields~$\phi$,
and $\vev{V\cdots V}$
are their correlation functions evaluated in the unperturbed vacuum.
For the even spin structures all vertices $V_\phi$ should be taken in
the zero-ghost-number picture;
in the odd (Ramond-Ramond) spin structure one of the vertices
should be taken in a different picture and an
additional operator should be inserted in the correlation function.
\Ref\FMS{D.~Friedan, E.~Martinec and S.~Shenker
 \journal Nucl. Phys. &B271 (86) 93.}
Fortunately, the only feature of the Ramond-Ramond one-loop
amplitudes
that is relevant in this article is that all such amplitudes
contain the $\epsilon^{\alpha\beta\gamma\delta}$ tensor.
The amplitude we would like to compute --- $F_{\mu\nu}^2$ term in
${\cal L}_1$ --- does not contain the $\epsilon$ tensor,
so it will get no contribution from the
Ramond-Ramond partition function $Z_{11}(F_{\mu\nu})$.

Before we plug a perturbation that turns on a uniform gauge
field and no other fields into \perturbation,
we have to verify that this perturbation
obeys the classical string equations of motion.
Perturbatively, those equations can be derived from an effective
Lagrangian,
\Ref\CMPF{C.~Callan, I.~Klebanov and M.~Perry
 \journal Nucl. Phys. &B278 (86) 78.}
and in the absence of charged background fields, Yang-Mills equations
become $\delta{\cal L}/\delta A^\mu =
D^\nu (\delta{\cal L}/\delta F^{\mu\nu})=0$;
such equations are always satisfied by covariantly constant
$F_{\mu\nu}$
regardless of details of ${\cal L}(F_{\mu\nu})$.
On the other hand, non-zero $F^a_{\mu\nu}$ results in non-zero
classical stress-energy tensor
$T_{\mu\nu}=(F^a_{\mu\lambda}F^{a\lambda}_\nu-\coeff14 g_{\mu\nu}
F^a_{\alpha\beta} F^{a\alpha\beta})/(g_a^{\rm tree})^2 +O(F^4)$.
Therefore, Einstein equations can only be satisfied if we turn on the
background gravitational field in addition to the gauge field,
and both kind of vertices should appear in
the perturbative expansion~\perturbation.
Terms that are quadratic in $F^a_{\mu\nu}$ can be written as:
$$
  \left.{Z(F)\over Z(0)}\right|_{F^2}\ =\,
  \int\!\!d^2\zeta\,\vev{V_{\rm grav}(\zeta,\bar\zeta)}\ +\
  \half\!\int\!\!\!\!\int\!\!d^2\zeta_1\,d^2\zeta_2
  \vev{V_{\rm gauge}(\zeta_1,\bar\zeta_1)\,
  V_{\rm gauge}(\zeta_2,\bar\zeta_2)} .
 \eqn\secondorder
$$

Let me consider the second term first.
For the background $F_{\mu\nu}=\rm const$,
$A_\mu(X)={-1\over2}F_{\mu\nu}X^\nu$,
a zero-ghost-number-picture vertex for the background gauge field
\Ref\Sen{A.~Sen\journal Phys. Rev. &D32 (85) 2102.}
looks like
$$
  V_{\rm gauge}(\zeta,\bar\zeta)\ =\
  {i\alpha'\over 4\pi}\,F_{\mu\nu}\,
  \Bigl( X^\mu(\zeta,\bar\zeta)\bar\partial\)X^\nu(\zeta,\bar\zeta)\,
  +\,\Psi^\mu(\bar\zeta)\)\Psi^\nu(\bar\zeta) \Bigr)
  \cdot J^a(\zeta)\,,
 \eqn\gaugevertex
$$
where $X^\mu$ are space-time coordinates, $\Psi^\mu$ are their
fermionic
partners and $J^a$ is a current in the Kac-Moody algebra that
generates
the gauge group~$a$ under consideration.
The correlation function of two such vertices can be written as
$$
 \displaylines{
  \vev{V_{\rm gauge}(\zeta,\bar\zeta)\,V_{\rm gauge}(0)}\ =
  {{\alpha'}^2\over 8\pi^2}\; F_{\mu\nu}^2 \cdot
  \vev{J^a(\zeta)\,J^a(0)}
  \times \hfill \eqname\twovertex \cr \hfill \times
  \left( \vev{\Psi(\bar\zeta)\,\Psi(0)}^2 \,-\,
  \vev{\bar\partial X(\zeta,\bar\zeta)\, X(0)}^2 \right) \cr}
$$
(making use of the translational symmetry of the torus,
I have set $\zeta_2=0$).
Now consider the Kac-Moody factor in \twovertex.
In a sector of the Hilbert space that has definite charge $Q_a$
with respect to the zero mode of $J^a$, one has
$$
  \vev{J^a(\zeta)\,J^a(0)}(\tau)\ =
  \sum_n {\pi^2\,k_a\over \sin^2(\pi\zeta+n\pi\tau)}\
  -\ 4\pi^2 Q_a^2\ ;
 \eqn\currents
$$
in a more general situation the $Q_a^2$ term should be averaged.
All other terms are proportional to the Kac-Moody central charge
$k_a$ and do not otherwise depend on the choice of a particular
gauge group~$a$.
This is a consequence of the requirements that this correlation
function
should be holomorphic, single-valued on the torus, and have a
$k_a/\zeta^2$ singularity for $\zeta\to0$;
these requirements completely determine $\vev{J^a(\zeta)J^a(0)}$
up to a constant with respect to $\zeta$.
Note that denoting the Kac-Moody central charges with $k_a$ is not
in collision with the earlier use of the same symbols in eq~\deltaeq:
it is these central charges that determine the ratios between the
tree-level gauge couplings. \refmark{\PG}
for

Evaluation of the gravitational term in \secondorder\ requires an
explicit expression for $V_{\rm grav}$ that corresponds to a solution
of the Einstein equations with the source~$T_{\mu\nu}\)$.
\vadjust{\penalty  1000}%
However, I do not have to solve those equations to notice that
\vadjust{\penalty 10000}%
$V_{\rm grav}\propto T_{\mu\nu}\propto (g_a^{\rm tree})^{-2}
\propto k_a$
\vadjust{\penalty -1000}%
and does not otherwise depend on the choice of the gauge group~$a$.
Therefore, at this point I can write the following formula for
$\ws_a$:
$$
 \displaylines{
  \ws_a\ =\ k_a\cdot Y\ -\
  \half{\alpha'}^2 \!\int_\Gamma\!\! d^2\tau \sum_{\rm even\,\bf s}
  \Tr_{s_1}\left( (-)^{s_2F}\,q^H\,\bar q^{\bar H}\cdot Q_a^2 \right)
  \times\hfill \eqname\intermediate \cr
  \hfill\times \int\!\! d^2\zeta\,\left(
  \vev{\Psi(\bar\zeta)\,\Psi(0)}^2_{\bf s} -
  \vev{\bar\partial X(\zeta,\bar\zeta)\,X(0)}^2 \right) ,\cr }
$$
where $Y$ stands for the total of $a$-independent terms.
Notice that the $Q_a^2$ operator has been brought inside the trace
--- that is what I meant by averaging the $Q_a^2$ term.

The trace in \intermediate\ obviously depends on a particular
string vacuum.
Nevertheless, all four-dimensional vacua share some of the
world-sheet
degrees of freedom, namely the four $X^\mu$, $\Psi^\mu$ and the
superconformal ghosts.
Hence, we can decompose the trace in \intermediate\ as
$$
  Z_{\rm 4d+gh}(\tau,\bar\tau,{\bf s})\cdot
  \Tr_{s_1}\left( Q_a^2\cdot (-)^{s_2 F}\, q^H\,
       {\bar q}^{\bar H} \right)_{\rm int\,,}
 \eqn\tracedef
$$
where the second factor involves the `internal' degrees of freedom
only.
The first factor ---
the partition function of space-time and ghost
degrees of freedom --- is common to all four-dimensional models.
The effect of ghosts is to cancel the contribution of the non-zero
modes of two of the four $X^\mu$ and two of $\Psi^\mu$;
the remaining factor is a minus sign for all spin structures
other than~(NS,NS).
The non-zero modes of the other two $X^\mu$ contribute
$Z_X(\tau,\bar\tau)=\left|\eta(\tau)\right|^{-4}$,
and the contribution of two $\Psi^\mu$ can also
be expressed in terms of $\eta$-functions:
$$
  Z_\Psi(\bar\tau,{\bf s})\ =\ {1\over \eta^2(\bar\tau)}\cdot
  \cases{ \eta^2({\bar\tau+1\over2}) & for $\bf s\rm=(NS,NS)$,\cr
          \eta^2({\bar\tau  \over2}) & for $\bf s\rm=(NS,R )$,\cr
         2\eta^2(2\bar\tau)          & for $\bf s\rm=(R ,NS)$.\cr }
 \eqn\zpsi
$$
Finally, the contribution of zero modes of $X^\mu$ ($\Psi^\mu$
have no zero modes for even spin structures) is simply
$$
  Z_0(\tau,\bar\tau)\ = \int\! {d^4 p\over(2\pi)^4}\;
  \exp\left(-\pi\alpha'\tau_2 p^2\right)\ =\
  (2\pi)^{-4}\,(\alpha'\tau_2)^{-2} .
 \eqn\zmodes
$$

The last factor in eq.~\intermediate\ --- the world sheet
integral --- is also common to all heterotic string models
and can be computed explicitly.
The details of this computation are rather boring, but the result
can be summarized in terms of~$Z_\Psi$:
$$
  \int\!{d^2\zeta\over\tau_2}\,
  \left( \vev{\Psi(\bar\zeta)\,\Psi(0)}_{\bf s}^2 -
  \vev{\bar\partial X(\zeta,\bar\zeta)\,X(0)}^2 \right) \ =\
  4\pi i\,{d\ \over d\bar\tau} \log Z_\Psi(\bar\tau,{\bf s})\,.
 \eqno\eq
$$
This completes the analysis of eq.~\intermediate,
which can now be rewritten as
$$
  \ws_a\ =\ {1\over 16\pi^2}\!\int_\Gamma \!{d^2\tau\over\tau_2}\;
  \B_a(\tau,\bar\tau)  \ +\ k_a\cdot Y ,
 \eqn\stringren
$$
where
$$
  \B_a(\tau,\bar\tau)\ =\ {2\over\left|\eta(\tau)\right|^4} \cdot
  \sum_{\rm even\,\bf s} (-)^{s_1+s_2}\,
  {d Z_\Psi(\bar\tau,{\bf s}) \over 2\pi i\, d\bar\tau}\cdot
  \Tr_{s_1}\left( Q_a^2\cdot (-)^{s_2 F} q^H
  {\bar q}^{\bar H} \right)_{\rm int\,.}
 \eqn\beq
$$

The integral in eq.~\stringren\ is logarithmically divergent in the
infrared $\tau_2\to\infty$ limit.
This is similar to the divergence of the proper-time integral in
eq.~\fieldren;
in fact, both divergences have the same coefficient $b_a\)$.
To see that, consider the limit of $\B_a(\tau_1,\tau_2)$ as
$\tau_2$ becomes very large.
Unless we are dealing with a string model that has charged tachyons
--- an obviously unrealistic case --- the traces in eq.~\beq\
are dominated in this limit by massless particles that have $H=\bar
H=0$.
Hamiltonians $H$ and $\bar H$ are totals of space-time, ghost and
internal components; evaluating the space-time and ghost components,
we find that massless vectors have $(H,\bar H)_{\rm
int}=(\coeff1{12},
-\coeff38)$, massless fermions have $(\coeff1{12},0)$ and
massless scalars have $(\coeff1{12},\coeff18)$.
As to the infrared limit of the $\eta$ and $Z_\psi$
factors in eq.~\beq,
we know that $\eta(\tau)\approx q^{1/24}(1-q+\cdots)$, and we can use
formulas \zpsi\ to derive
$Z_\Psi(\bar\tau,{\rm NS})\approx \bar q^{-1/24} \pm 2\bar
q^{+11/24})$
and $Z_\Psi(\bar\tau,{\rm R})\approx 2\bar q^{+1/12}$.
Substituting this asymptotic behavior into eq.~\beq, we arrive at
$$
  \B_a\,\becomes{\tau_2\to\infty}\,
  -\coeff{11}{3}\tr_{V,M=0}(Q_a^2) \,+\, \coeff23\tr_{F,M=0}(Q_a^2)
  \,+\, \coeff13\tr_{S,M=0}(Q_a^2)\
  =\ b_a \equiv \lim_{t\to\infty} \b_a(t)
 \eqn\irlimit
$$
(\cf~eq.~\irfield):
the string theory indeed yields the same infrared divergence
of ${\cal W}_a$ as the effective low-energy gauge theory.
(~Notice that the width of $\Gamma$ ---\penalty-1000{}
$w_\Gamma(\tau_2)\equiv\int\!d\tau_1\)\theta(\tau_1+i\tau_2\in\Gamma)
$
--- is 1 for $\tau_2\ge1$.)
Similar to the GUT case, this means that despite the divergence
of $\ws_a$,
the difference between the bare couplings comes out finite.
Let me write down an explicit formula:
\vadjust{\vfootnote\footsymbolgen {
  To see that $t$ should be identified with $\pi\alpha'\tau_2\)$,
  compare the exponential factors associated with the space-time
  momentum.
  For the proper time $t$ this factor is $e^{-tp^2}$
  (\cf~eq.~\propertimeq);
  the corresponding factor for $\tau_2$ is $(q\bar q)^{\alpha'p^2/4}=
  e^{-\pi\tau_2\alpha'p^2}$ (\cf~eq.~\zmodes).}
 \nobreak }
$$
\displaylines{
{16\pi^2\over g_a^2}\,-\,{16\pi^2 k_a\over \gstr^2}\
    = \int_\Gamma\!{d^2\tau\over\tau_2}\,
    \bigl( \B_a(\tau_1,\tau_2) - b_a \bigr)\
    \hfill \eqname\ginterm \cr
\hfill  {}+\ b_a\cdot\!\int_0^\infty\!{d\tau_2\over \tau_2}\,
    \bigl( w_\Gamma(\tau_2) - C_\Lambda(\pi\alpha'\tau_2)^\footsymbol
    \bigr)\ +\ k_a\cdot 16\pi^2 Y .\cr }
$$

It remains to translate the bare couplings into renormalized
couplings
evaluated at $\mu=\mgut$;
this is a straightforward exercise for the field-theory
coupling~$g_a\)$, but how
does one chose a renormalization scheme for the string coupling?
One logical choice is to use the ultraviolet finiteness of the
string theory and to define the renormalized $\gstr$ to be
the same as the bare $\gstr\)$.
However, I prefer a different scheme for the $\ggut\)$, namely
$\ggut^{-2}\eqdef (\gstr^{\rm bare})^{-2}+Y$;
in this scheme the last term in \ginterm\ is precisely canceled,
so I do not even have to compute the~$Y$
(and I didn't).
Of course, I will have to compute the $Y$ if I ever need a value
of $\gstr^{-2}$ that is accurate to the order $O(1)$,
but I do not need it in this article.

Finally, when I translate the bare $g_a$ into the $\msb$ coupling
$g_a(\mu=\mgut)$, I add a
$b_a\cdot(\xi-\log{\Lambda^2\over\mgut^2})$
term to the left hand side of eq.~\ginterm;
$\xi$ depends on a choice of the regularization scheme.
The second term on the right hand side of \ginterm\ can be written
in a similar form: $b_a\cdot(\xi'-\log\alpha'\Lambda^2)$;
the difference between these two terms is
$b_a\cdot(\xi''-\log\alpha'\mgut^2)$ where $\xi''=\xi'-\xi$
is a numerical constant.
A straightforward exercise in calculus yields
$\xi''=1+\log(2/\sqrt{27}\pi)-\gamma\approx-1.6767$\penalty-300\
(${\gamma\approx 0.57722}$ is the Euler's constant).
Now consider the fact that a precise definition of $\mgut$ in
terms of heavy masses is a matter of convention.
No such convention has been established for the string-based models,
so I consider myself free to define
$\mgut\eqdef e^{\xi''/2}/\sqrt{\alpha'}$;
the virtue of this definition is that it reduces the formula for the
threshold corrections $\Delta_a$ to the first term in eq.~\ginterm.

At this point I can write down
{\bf the main result of this article:
\it one-string-loop threshold corrections to low-energy gauge
couplings
are related to the model's spectrum via:}
$$
  \Delta_a\ = \int_\Gamma\!{d^2\tau\over\tau_2}\,
  \left( \B_a(\tau,\bar\tau)\,-\,b_a \right) ,
 \eqn\mainresult
$$
{\it where $b_a=\B_a(\tau=i\infty)$ and
$\B(\tau,\bar\tau)$ is defined in eq.~\beq.}
This result is valid in the $\msb$ renormalization scheme for
the following
{\it definition of the unification scale:}%
\vadjust{\vfootnote\footsymbolgen{%
    I am using the following conventions for the couplings:
    $\ggut$ and the individual gauge couplings $g_a$ are normalized
    according to the conventions used in the standard model
phenomenology
    and ordinary GUTs.
    The string coupling $\gstr$ is normalized according to the
    tree-level relation
    $\ggut\buildrel\rm tree\over{\Relbar\joinrel\Relbar}\gstr$.
    With this convention, $\alpha'\mpl^2\gstr^2=32\pi$,
    which differs by a factor of 2 from a similar formula
    given in ref.~[\PG].}
  \global\let\firstsymbol=\footsymbol
  \vfootnote\footsymbolgen{%
    The last two digits here should not be taken too seriously
    since this formula is only accurate in its leading term.}}
$$
\eqalign{
\mgut\ \eqdef\ {2e^{(1-\gamma)/2}\, 3^{-3/4}\over\sqrt{2\pi\alpha'}}\
&
=\ {e^{(1-\gamma)/2}\, 3^{-3/4}\over 4\pi}\; \gstr \mpl^\firstsymbol
\cr
&\approx\ \ggut\cdot 5.27\cdot 10^{17}\,\hbox{GeV.}^\footsymbol \cr
}\eqn\gutdef
$$

I would like to stress that formula~\mainresult\ is valid for {\sl
any}
four-dimensional string model that has no tachyons and is based upon
a
{\sl classical vacuum of the heterotic string},
\ie, a world-sheet field theory with exact
$(0,1)$ superconformal invariance. \refmark{\FMS}
No additional requirements are necessary;
in particular, a model need not be space-time supersymmetric.

\chapter{Threshold Effects in Orbifold Models}

Formulae \mainresult\ and \beq\ express the one-loop threshold
corrections in terms of the entire particle spectrum of the model
under consideration.
Unfortunately, for many string models such as
Calabi-Yau compactifications of the ten-dimensional heterotic string
\Ref\CHSW{P.~Candelas, G.~Horowitz,
 A.~Strominger and E.~Witten \journal Nucl.\ Phys. & B258 (85) 46.}
only the massless spectra are known.
On the other hand, for orbifolds and other models built out of
lattices, free fermions, Kac-Moody algebras, minimal Virasoro
algebras,
\etc, writing down the entire spectrum --- massless and massive ---
and computing the traces in formula~\beq\ is a matter of algebra.
In particular, for an orbifold with point group~$G$, the traces
can be expanded as
$$
  \Tr_{s_1}\left( (-)^{s_2 F} q^H {\bar q}^{\bar H} \cdot Q_a^2
\right)
  \ =\ {1\over|G|} \sum_{g,h\in G \atop gh=hg}
  \Tr_{(g,s_1)} \left( h\cdot
   (-)^{s_2 F} q^H {\bar q}^{\bar H} \cdot Q_a^2 \right) ,
 \eqn\orbitrace
$$
and all traces on the right hand side of \orbitrace\ can be easily
written in terms of $\eta$-functions and $\Theta$-functions of
various lattices.
Actually, the completely untwisted sector ($g=h=1$) does not
contribute
anything to the $\B_a(\tau,\bar\tau)$, because this sector by itself
is
an $N=4$ supersymmetric model.
In addition to $\Psi^\mu$ it has six more free world-sheet fermions
that obey the same boundary conditions while no other world-sheet
fields are affected by a choice of a spin structure~$\bf s$.
Therefore, the relevant traces over the internal degrees of freedom
can be written as $Z_\Psi^3({\bf s},\bar\tau)$ times some trace
that does not depend on~$\bf s$.
Hence, in the $N=4$ supersymmetric case the right hand side of \beq\
is proportional to
$$
  \sum_{\rm even\ \bf s} (-)^{s_1+s_2}\, Z_\Psi^3({\bf s},\bar\tau)
  \cdot {d\over d\bar\tau} Z_\Psi({\bf s},\bar\tau)\ =\
  \coeff14\){d\over d\bar\tau} \sum_{\rm even\ \bf s} (-)^{s_1+s_2}\,
  Z_\Psi^4({\bf s},\bar\tau)\ =\ 0.
$$
This means that
{\sl for $N=4$ supersymmetric models not only the $\beta$-functions
vanish, but the one-loop threshold corrections vanish too.}

I would like to conclude this article with an example of actually
computing the threshold effects for some simple model.
Unfortunately, no realistic or almost-realistic models fall into
this category, so I shall use an unrealistic example of the $Z_3$
orbifold.
\Ref\DHVW{L.~J.~Dixon, J.~A.~Harvey, C.~Vafa and E.~Witten
 \journal Nucl.\ Phys. & B261 (85) 671 and {\bf B274} (1986), 285.}
The symmetric $Z_3$ orbifold
is an $N=1$ supersymmetric model whose low-energy gauge group
is $SU(3)\otimes E_6\otimes E_8$;
charged massless matter fields transform as
$3({\bf3},{\bf27},{\bf1})+27({\bf1},{\bf27},{\bf1})+
81({\bf\bar3},{\bf1},{\bf1})$.
The $\beta$-functions are given by $b_8={-90}$, $b_6=b_3={+72}$,
so the difference between low-energy couplings for the $E_6$ and
for the $SU(3)$ is due solely to the threshold corrections.
Actually, the term `symmetric $Z_3$ orbifold' refers to a nine-%
parameter family of string models, where the parameters are the
radii and the angles of a torus that is modded out by the $Z_3$
group.
However, out of the nine $(g,h)$ sectors whose contributions total
the $\B_a\)$, the only sector affected by the geometry of the torus
is the completely untwisted sector $g=h=1$, and that sector is
$N=4$ supersymmetric and does not contribute anything at all.
Hence, none of the moduli of the $Z_3$ orbifold affect
the values of the threshold corrections.
This feature is probably peculiar to the $Z_3$ case: most other
orbifolds contain $N=2$ supersymmetric sectors whose contributions
do depend on some of the moduli.

As to the actual calculation, it is not too hard to derive algebraic
expressions for $\B_{3,6,8}(\tau,\bar\tau)$ in terms of $\eta(\tau)$
and the $\Theta$-functions of the $SU(3)$ lattice;
these expressions are rather long and not particularly interesting.
The only interesting feature is that $\B_3$ and $\B_6$ are exactly
equal to each other, so there is no one-loop difference between $g_3$
and $g_6\)$.
$\B_8$ is different; integrating the difference numerically,
I got $\Delta_3-\Delta_8\approx11$.
When compared to $b_3-b_8=162$, this threshold correction
amounts to a small change in the effective GUT scale (defined as a
point at which the renormalization curves for $g_3(\mu)$ and
$g_8(\mu)$
cross each other): it comes out about 3.5\% higher than \gutdef.

Surprisingly, I got exactly the same answer for a different $Z_3$
orbifold model in which each of the two $E_8$ groups is broken down
to an $E_6\otimes SU(3)$.
For this model $b_3-b_6=81=162/2$, and the difference $\B_3-\B_6$ is
also
exactly one half of the expression for $\B_3-\B_8$ in the first
example,
so the effective GUT scale is the same in both models.
Moreover, if the moduli of the orbifold are chosen such as to yield
three extra $SU(3)$ factors in the low-energy gauge group,
the running coupling $g'_3(\mu)$ for those factors  crosses both
$g_3(\mu)$ and $g_6(\mu)$ at the same point.
It is not clear at the moment whether these coincidences are peculiar
to $Z_3$ orbifolds, or whether they are common to other string
models.
In the latter case this would be a yet another class of string
miracles;
further investigation of this subject might be worthwhile.

\smallskip\noindent
The author would like to {\bf acknowledge} many fruitful
conversations
with T.~Banks, M.~Peskin, L.~Susskind and other members and guests
of the SLAC theory group.

\refout
\bye